\documentstyle[11pt,newpasp,twoside]{article}
\def\HST{{\it HST}}

\def\ASCA{{\it ASCA}}

\def\Chandra{{\it Chandra}}
\def\FUSE{{\it FUSE}}
\def\arcsec{\ifmmode '' \else $''$\fi}
\def\arcmin{\ifmmode ' \else $'$\fi}
\def\arcsecpoint{\ifmmode ''\!. \else $''\!.$\fi}
\def\arcminpoint{\ifmmode '\!. \else $'\!.$\fi}
\def\cc{\ifmmode {\rm cm}^{-3} \else cm$^{-3}$\fi}
\def\cl{\ifmmode {\rm cm}^{-2} \else cm$^{-2}$\fi}
\def\micron{\ifmmode \mu{\rm m} \else $\mu$m\fi}
\def\kms{\ifmmode {\rm km\,s}^{-1} \else km\,s$^{-1}$\fi}
\def\Hubble{\ifmmode {\rm km\,s}^{-1}\,{\rm Mpc}^{-1}
        \else km\,s$^{-1}$\,Mpc$^{-1}$\fi}
\def\ergsec{\ifmmode {\rm ergs\;s}^{-1} \else ergs s$^{-1}$\fi}
\def\ergscm{\ifmmode {\rm ergs\,s}^{-1}\,{\rm cm}^{-2}
          \else ergs\,s$^{-1}$\,cm$^{-2}$\fi}
\def\ergscmA{\ifmmode {\rm ergs\,s}^{-1}\,{\rm cm}^{-2}\,{\rm \AA}^{-1}
          \else ergs\,s$^{-1}$\,cm$^{-2}$\,\AA$^{-1}$\fi}
\def\ergscmHz{\ifmmode {\rm ergs\,s}^{-1}\,{\rm cm}^{-2}\,{\rm Hz}^{-1}
          \else ergs\,s$^{-1}$\,cm$^{-2}$\,Hz$^{-1}$\fi}
\def\Msun{\ifmmode M_{\odot} \else $M_{\odot}$\fi}
\def\Lsun{\ifmmode L_{\odot} \else $L_{\odot}$\fi}
\def\qo{\ifmmode q_{0} \else $q_{0}$\fi}
\def\Ho{\ifmmode H_{0} \else $H_{0}$\fi}
\def\lya{L$\alpha$}

\def\civ{C\,{\sc iv}}

\newcommand{\ovi}{O~{\sc vi}}

\def\edcomment#1{\iffalse\marginpar{\raggedright\sl#1\/}\else\relax\fi}
\reversemarginpar

\begin{document}
\title{FUSE Observations of Warm Absorbers in AGN}
\author{Gerard A. Kriss}
\affil{Space Telescope Science Institute, 3800 San Martin Drive,
Baltimore, MD 21218}

\begin{abstract}
In a survey of the UV-brightest AGN using the {\it Far Ultraviolet Spectroscopic
Explorer (FUSE)}, we commonly find associated absorption in the
{\sc O~vi} $\lambda\lambda1032,1038$ resonance doublet.
Of 34 Type I AGN observed to date with $z < 0.15$,
16 show detectable {\sc O~vi} absorption.
Most absorption systems show multiple components with intrinsic widths of
$\sim100~\kms$ spread over a blue-shifted velocity range of $< 1000~\kms$.
With the exception of three galaxies (Ton~S180, Mrk~478, and Mrk~279),
those galaxies in our sample with existing X-ray or longer wavelength UV
observations also show {\sc C~iv} absorption and evidence of a soft X-ray
warm absorber.
In some cases, a UV absorption component has physical properties
similar to the X-ray absorbing gas, but in others there is no clear
physical correspondence between the UV and X-ray absorbing components.
\end{abstract}

\section{Introduction}

Roughly 50\% of all Seyfert galaxies show UV absorption lines, most
commonly seen in {\sc C~iv} and Ly$\alpha$ (\cite{Crenshaw99}).
X-ray ``warm absorbers" are equally common in Seyferts
(\cite{Reynolds97}; \cite{George98}).
Crenshaw et al. (1999) note that all instances of X-ray absorption
also exhibit UV absorption.
While Mathur et al. (1994; 1995) have suggested that the same
gas gives rise to both the X-ray and UV absorption, the spectral complexity
of the UV and X-ray absorbers indicates that a wide range of physical
conditions are present.
Multiple kinematic components with differing physical conditions are seen
in both the UV (\cite{Crenshaw99}; \cite{Kriss00b}) and in the X-ray
(\cite{Kriss96a}; \cite{Reynolds97}; \cite{Kaspi01}).

The short wavelength response (912--1187 \AA) of the
{\it Far Ultraviolet Spectroscopic Explorer (FUSE)}
(\cite{Moos00}; \cite{Sahnow00})
enables us to make high-resolution spectral measurements
($R \sim 20,000$) of the high-ionization
ion {\sc O~vi} and the high-order Lyman lines of neutral hydrogen.
The \ovi\ doublet is a crucial link for establishing a connection between
the higher ionization absorption edges seen in the X-ray and the lower
ionization absorption lines seen in earlier UV observations.
The high-order Lyman lines provide a better constraint on the
total neutral hydrogen column density than Ly$\alpha$ alone.
Lower ionization species such as {\sc C~iii} and {\sc N~iii} also have strong
resonance lines in the \FUSE\ band, and these often are useful for setting
constraints on the ionization level of any detected absorption.
The Lyman and Werner bands of molecular hydrogen also fall in the \FUSE\ band,
and we have serached for intrinsic $\rm H_2$ absorption that may be
associated with the obscuring torus.

We have been conducting a survey of the $\sim80$ brightest AGN using
{\it FUSE}. To date (March 1, 2001) we have observed a total of 57;
of these, 34 have $z < 0.15$, so that the {\sc O~vi} doublet is visible
in the FUSE band.  In this presentation, I will be talking about the UV
absorption properties of this sub-sample.
A more extensive discussion of the full survey can be found
in my review from last summer (\cite{Kriss00c}).
Results on \FUSE\ observations of individual interesting objects such as
NGC~3516 (Hutchings et al.  2001), NGC~3783 (Kaiser et al. 2001), and
NGC~5548 (Brotherton et al. 2001) can also be found in these proceedings.

\section{Survey Results}

\begin{figure}[b]
\plotfiddle{"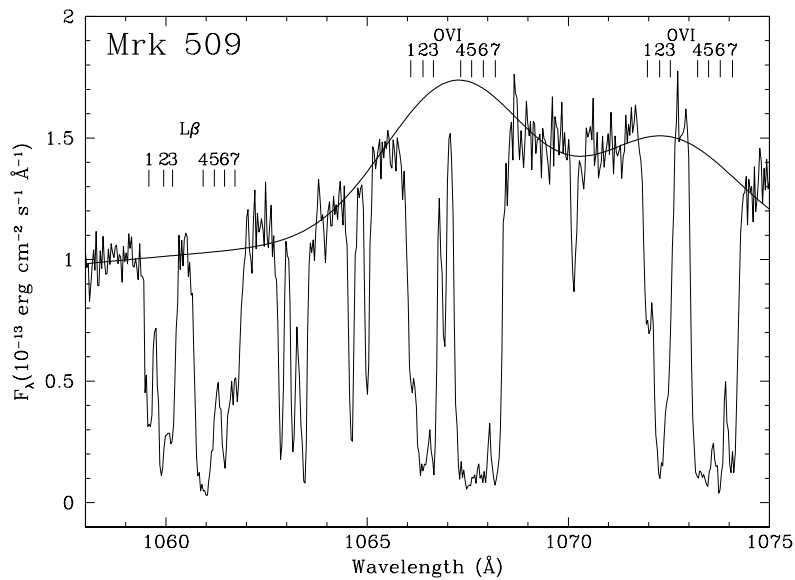"}{2.8in}{-90}{40}{40}{-157}{225}
\caption{
The Ly$\beta$/O~VI spectral region of Mrk~509, adapted from Kriss et al.
(2000), illustrates the ``blend" spectral morphology discussed in the text.
The smooth, heavy line shows the local continuum
comprised of a powerlaw continuum plus O~VI emission from Mrk~509.
Seven blended kinematic components are marked.
}
\end{figure}

Roughly 50\% (16 of 34) of the low-redshift AGN observed using \FUSE\ show
detectable \ovi\ absorption.  None show $\rm H_2$ absorption.
I'd first like to review the spectral morphology of the \ovi\ absorption
features.
We see three basic morphologies:
(1) {\bf blend}: multiple \ovi\ absorption components that are blended
together.  This is the most common morphology
(8 of 16 objects fall in this class), and the spectrum of Mrk~509
shown in Figure~1 illustrates the typical appearance.
(2) {\bf single}: single, narrow, isolated \ovi\ absorption lines, as
illustrated by the spectrum of Ton~S180 in Figure 2.
(3) {\bf smooth}: this is an extreme expression of the ``blend" class,
where the \ovi\ absorption is so broad and blended that individual
\ovi\ components cannot be identified.  The spectrum of NGC~4151 shown in
Figure~3 is typical of this class. (Note, however, that both broad smooth
absorption as well as discrete components are often visible in this class.)

As shown in the summary of characteristics presented in Table~1,
individual \ovi\ absorption components have 
FWHM of 50--750 \kms, with most objects having FWHM $< 100~\kms$.
The multiple components that are typically present are almost always
blue shifted, and they span a velocity range of 200--4000 \kms;
half the objects span a range of $< 1000~\kms$.

\begin{table}
\caption{Properties of O VI Absorbers in AGN Observed with FUSE}
\begin{tabular*}{5.2in}{l c c c c c c}
\tableline
Name & \# Comp & Type & Line   & Velocity & Other   & X-ray  \\
     &         &      & Widths & Spread   & $\rm~UV~Abs.^a$ & $\rm~Abs.^b$  \\
     &         &      & ($\rm km~s^{-1}$) & ($\rm km~s^{-1}$) & & \\
\tableline
I Zw 1      & 2 & single &  50 &  850 & Y & \ldots \\
NGC 985     & 1 & smooth & 100 & 1000 & \ldots & \ldots \\
NGC 3516    & 5 & blend  & 100 & 1000 & Y & Y \\
NGC 3783    & 4 & blend  & 300 & 1600 & Y & Y \\
NGC 4151    & 3 & smooth & 100 & 1600 & Y & Y \\
NGC 5548    & 6 & blend  &  50 & 1300 & Y & Y \\
NGC 7469    & 3 & single &  50 & 1000 & Y & Y \\
Mrk 279     & 5 & blend  & 100 & 1600 & Y & N \\
Mrk 290     & 1 & single & 200 &  400 & \ldots & Y \\
Mrk 304     & 5 & smooth & 750 & 1500 & \ldots & \ldots \\
Mrk 478     & 5 & blend  &  50 & 2700 & N & \ldots \\
Mrk 509     & 5 & blend  &  50 &  700 & Y & Y \\
Mrk 817     & 2 & blend  & 250 & 4000 & \ldots & \ldots \\
PG1351+64   & 4 & blend  & 100 & 2000 & Y & \ldots \\
Ton 951     & 1 & single &  40 &  200 & \ldots & \ldots \\
Ton S180    & 3 & single &  50 &  300 & N & N \\
\tableline
\tableline
\end{tabular*}
\parbox{5.85in}{
$\rm ^a$UV reference: Crenshaw et al. (1999).\\
$\rm ^b$X-ray references: George et al. (1998); Reynolds (1997).
}
\end{table}

The 50\% \ovi\ absorption fraction is comparable to those Seyferts that show
longer-wavelength UV (\cite{Crenshaw99})
or X-ray (\cite{Reynolds97}; \cite{George98}) absorption.
In Table~1 we make a detailed comparison to these UV and X-ray absorption
studies.
We can summarize this comparison in two ways:
(1) all objects that have shown previous evidence of either UV or X-ray
absorption also show \ovi\ absorption in our \FUSE\ observations;
(2) of those objects showing detectable \ovi\ absoption that also have
previous UV or X-ray observations available, only 3 do not show X-ray
absorption or longer-wavelength UV absorption.
These three exceptions deserve some attention:

{\bf Ton S180} was observed simultaneously with \FUSE, \Chandra, and \HST\ 
(\cite{Turner01}).
As shown in Figure 2, the absorption features observed with \FUSE\ are
quite weak.  At the resolution of the \HST\ observations, which used the
STIS low-resolution gratings, corresponding \civ\ and \lya\ absorption
would not be expected to be detectable.
No X-ray absorption features are seen in the \Chandra\ grating observation,
which implies that either the total column density of the absorbing gas is
very low, and/or that its ionization state is lower than that necessary to
create X-ray absorbing species.

{\bf Mrk 478} was observed previously with \HST\ using the FOS.
As for Ton~S180, these observations were low spectral resolution, so that
weak \civ\ or \lya\ absorption at the equivalent widths observed for \ovi\ 
would not be detectable.

{\bf Mrk 279} shows strong Ly$\alpha$ absorption in an archival \HST\ spectrum.
However, the \ASCA\ observation of Mrk~279 does not show evidence of a warm
absorber (\cite{Weaver01}).  This was a short observation, however, and
total column densities of $10^{21}~\rm cm^{-2}$ could easily be present,
which would be detectable in \Chandra\ grating observations.

\begin{figure}
\plotfiddle{"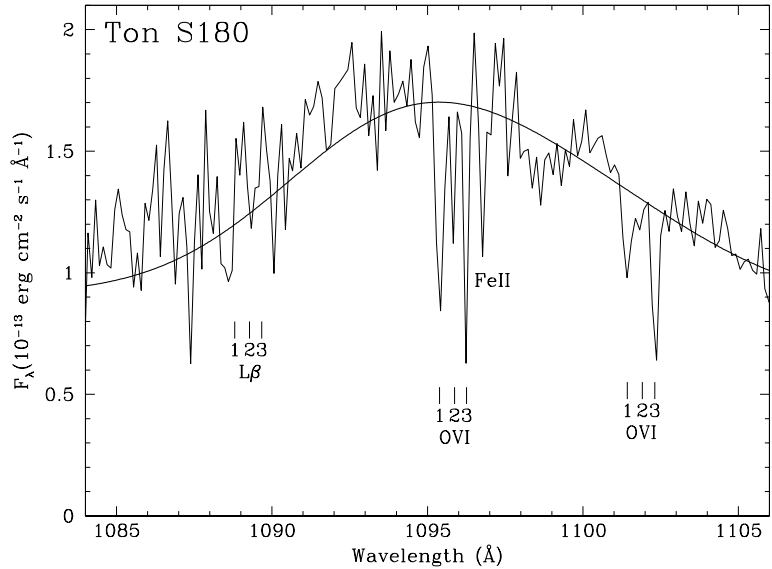"}{2.8in}{-90}{40}{40}{-157}{225}
\caption{
The  Ly$\beta$/O~VI spectral region of Ton~S180 shows the ``single"
absorption line spectral morphology discussed in the text.
As in Fig. 1, the heavy smooth line shows the sum of a powerlaw continuum
plus broad O~VI emission.
Three, narrow, isolated absorption components are visible in O~VI.
These are marked for both O~VI and Ly$\beta$, even though no absorption
is detected in Ly$\beta$.  Ton~S180 is also one of the objects in our
sample for which we do not detect longer wavelength UV absorption or
X-ray absorption (Turner et al. 2001).
}
\end{figure}

\begin{figure}
\plotfiddle{"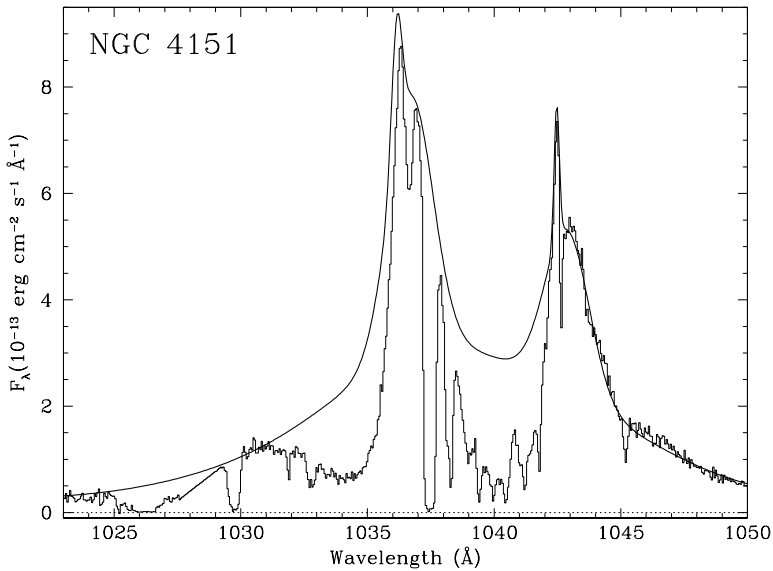"}{2.8in}{-90}{40}{40}{-157}{225}
\caption{
Our FUSE observation of NGC~4151 caught this galaxy in a low-luminosity state.
The O~VI region is dominated by narrow-line emission, but smooth, broad
absorption illustrating the ``smooth" morphology discussed in the text.
The heavy, smooth solid line shows the modeled total emission from a powerlaw
continuum plus weak broad-line and strong narrow-line emission.
}
\end{figure}

\section{Discussion}

The multiple kinematic components frequently seen in the UV absorption spectra
of AGN clearly show that the absorbing medium is complex, with separate
UV and X-ray dominant zones.
In some cases, the UV absorption component corresponding to the X-ray warm
absorber can be clearly identified (e.g., Mrk~509, Kriss et al. 2000).
In others, however, {\it no} UV absorption component shows physical
conditions characteristic of those seen in the X-ray absorber
(NGC~3783: Kaiser et al. 2001; NGC~5548: Brotherton et al. 2001).
One potential geometry for this complex absorbing structure is high density,
low column UV-absorbing clouds embedded in a low density,
high ionization medium that dominates the X-ray absorption.
As discussed by Krolik \& Kriss (1995; 2001), this is possibly a wind driven
off the obscuring torus.
At the critical ionization parameter for evaporation, there is a broad range of
temperatures that can coexist in equilibrium at nearly constant pressure;
for this reason, the flow is expected to be strongly inhomogeneous.
What would this look like in reality?
We will not soon get a close-up look at this aspect of an AGN, so it is
instructive to look at nearby analogies.
The HST images of the pillars of gas in the Eagle Nebula, M16,
show the wealth of detailed structure in gas evaporated from a molecular
cloud by the UV radiation of nearby newly formed stars (Hester et al. 1996).
In an AGN one might expect a dense molecular torus to be
surrounded by blobs, wisps, and filaments of gas at various densities.
It is plausible that the multiple UV absorption lines seen in AGN with
warm absorbers are caused by high-density blobs of gas embedded in a hotter,
more tenuous, surrounding medium, which is itself responsible for the X-ray
absorption.  Higher density blobs would have lower ionization
parameters, and their small size would account for the low overall column
densities.

In summary, we find that \ovi\ absorption is common in low-redshift ($z < 0.15$)
AGN. 16 of 34 AGN observed using \FUSE\ show 
multiple, blended \ovi\ absorption lines with typical widths
of $\sim 100~\kms$ that are blueshifted over a velocity range of $\sim$
1000 \kms.  With three exceptions,
those galaxies in our sample with existing X-ray or longer wavelength UV
observations also show {\sc C~iv} absorption and evidence of a soft X-ray
warm absorber.
In some cases, a UV absorption component has physical properties
similar to the X-ray absorbing gas, but in others there is no clear
physical correspondence between the UV and X-ray absorbing components.

\end{document}